\documentclass[twocolumn]{aastex631}

\usepackage{amsmath}

\begin{document}

\title{Exoplanet Ephemerides Change Observations (ExoEcho). II. Transit timing variation analysis of Brown Dwarfs around Solar-type Stars}

\correspondingauthor{Bo Ma}
\email{mabo8@mail.sysu.edu.cn}

\author{Wenqin Wang}
\affil{School of Physics and Astronomy, Sun Yat-sen University, Zhuhai 519082, China; {\it mabo8@mail.sysu.edu.cn}}
\affil{CSST Science Center for the Guangdong-Hong Kong-Macau Great Bay Area, Sun Yat-sen University, Zhuhai 519082, China}

\author{Xinyue Ma}
\affil{School of Physics and Astronomy, Sun Yat-sen University, Zhuhai 519082, China; {\it mabo8@mail.sysu.edu.cn}}
\affil{CSST Science Center for the Guangdong-Hong Kong-Macau Great Bay Area, Sun Yat-sen University, Zhuhai 519082, China}

\author{Zhangliang Chen}
\affil{School of Physics and Astronomy, Sun Yat-sen University, Zhuhai 519082, China; {\it mabo8@mail.sysu.edu.cn}}
\affil{CSST Science Center for the Guangdong-Hong Kong-Macau Great Bay Area, Sun Yat-sen University, Zhuhai 519082, China}

\author{Cong Yu}
\affil{School of Physics and Astronomy, Sun Yat-sen University, Zhuhai 519082, China; {\it mabo8@mail.sysu.edu.cn}}
\affil{CSST Science Center for the Guangdong-Hong Kong-Macau Great Bay Area, Sun Yat-sen University, Zhuhai 519082, China}

\author{Shangfei Liu}
\affil{School of Physics and Astronomy, Sun Yat-sen University, Zhuhai 519082, China; {\it mabo8@mail.sysu.edu.cn}}
\affil{CSST Science Center for the Guangdong-Hong Kong-Macau Great Bay Area, Sun Yat-sen University, Zhuhai 519082, China}

\author{Bo Ma}
\affil{School of Physics and Astronomy, Sun Yat-sen University, Zhuhai 519082, China; {\it mabo8@mail.sysu.edu.cn}}
\affil{CSST Science Center for the Guangdong-Hong Kong-Macau Great Bay Area, Sun Yat-sen University, Zhuhai 519082, China}

\begin{abstract}
Transit timing variation (TTV) is a useful tool for studying the orbital properties of transiting objects. 
However, few TTV studies have been done on transiting brown dwarfs (BDs) around solar-type stars.
Here we study the long-term TTV of a population of close BD companions around solar-type stars using TESS data. 
We use the measured orbital period change rate to constrain the tidal interaction strength between the host star and the BD companion and put limits on the destruction timescale of these transiting BDs. 
However, we find no statistically significant evidence of orbital decay or expansion in our sample based on the current data.
This may be due to either poor observational data or inherently weak tidal dissipation.
We then perform simulations to investigate future observation strategies for detecting orbital decay of transiting BDs, which show NGTS-7A b, TOI-263~b 
and LP 261-75 b are the most promising targets in the next few years.
Our study demonstrates the potential of TTV technique to probe the formation and evolution of close BD companions around solar-type stars.
\end{abstract}


\section{Introduction} \label{sec:intro}
Brown dwarfs (BDs), with masses ranging from approximately 13 to 80~$M_{\rm J}$, are celestial objects that occupy the middle ground between the largest gas giant planets and the smallest stars \citep{Basri2000, Burrows01}.
They are often called `failed stars' because their masses are insufficient to sustain hydrogen fusion in their cores. 
BDs may form through the fragmentation of molecular clouds, similar to the process by which stars form. 
These clouds contract under their own gravity, and as they do so, they can break up into smaller clumps, some of which may become BDs. 
There is also evidence to suggest that some BDs may form like planets, through the gradual accumulation of gas and dust in a protoplanetary disk, but with a mass that exceeds the traditional definition of a planet.

A significant scarcity of close BDs has been noted orbiting main-sequence (MS) solar-type stars, a phenomenon referred to as the `brown dwarf desert' \citep[BDD;][]{Halbwach00,Marcy00}.
\citet{Farihi05} proposed that white dwarf (WD) systems might also show a comparable lack of BD companions \citep[see also][]{Chen24}. 
The origin of this desert is thought of being associated with the distinct formation processes of giant planets and low-mass stars \citep{Burrows01, Grether06, Ma14, Persson19, Kiefer21, Feng22, Stevenson23}.
Therefore, examining the properties of BDs in the BDD can provide information valuable for the understanding of the planetary and star formation mechanisms around MS stars \citep{Ma14, Shahaf19}. 

One potential way to study the evolution of BD companions around solar-type stars, and the formation of the BDD, is to measure their orbital period decay.
Measuring the transit timing variation (TTV) of transiting exoplanets has become an effective technique to probe the dynamical properties of exoplanetary systems.
For example, the orbital decay of hot Jupiters (HJs) can be revealed through monitoring their transit timings over decades \citep{Ivshina22, Wang24}. 
For example, the detection of orbital period decay has been confirmed in the WASP-12 system through various TTV studies \citep{Maciejewski16, Patra17, Yee20}. 
Another orbital decay candidate is WASP-4~b \citep{Bouma19}, although the evidence is less compelling.
Short-term TTVs can also be used to detect extra planets in the same planetary system \citep{Holman05, Agol05}. 

\citet{Bowler20} argued that high-mass BDs predominantly form similarly to stellar binaries, based on a comparison of the mass-eccentricity distributions of BDs and giant planets. 
Since TTV studies can help reveal the dynamic properties of BD systems, expanding TTV analysis to transiting BDs can provide additional observational constraints on their formation and evolution theories. 
The Transiting Exoplanet Survey Satellite (TESS; \citealt{Ricker14}), launched in 2018, has provided such a great opportunity. 
It offers the latest and precise transit timing measurements for a lot of known transit BD companions, which are suitable for long-term TTV studies. 
Here we propose to study the orbital decay of BD companions around solar-type stars using TTV technique, as part of our ExoEcho project. The ExoEcho (Exoplanet Ephemerides CHange Observation) project is designed to study the photodynamics of exoplanets by leveraging high-precision transit timing data from ground- and space-based telescopes \citep{Wang24, Zhang24, Ma25}. 
By combining the high-precision 2-minute cadence transit data provided by TESS with archival data from previous literature, new constraints can be placed on the period change rate of these transiting BDs and the tidal dissipation factor of their host stars \citep{Wang24}.

The paper is organized as follows. Section 2 describes our sample selection and timing analysis. Section 3 introduces the transit timing models and the model fitting processes. Section 4 presents our results. In Section 5, we summarize our main findings and discuss the implications of our findings. 

\section{Data and Observation \label{sec:TESS} }
\subsection{Sample selection}
We focus on BDs that are likely to exhibit measurable orbital period changes over short timescales. The samples selected for this work are short-period transiting BDs with orbital periods $\lesssim$ 5 days and masses between 13 and 80 $M_J$, all of which have been observed by TESS.  
A total of 10 BD systems were analyzed: AD~3116~b, KELT-1~b, LP~261-75~b, WASP-30~b, WASP-128~b, EPIC~212036875~b, GPX-1~b, NGTS-7A~b, TOI-263~b, and TOI-503~b, with TOI-263 b and TOI-503 b were discovered by TESS itself.

\subsection{TESS data}
We downloaded the Presearch Data Conditioning-Simple Aperture Photometry (PDC-SAP) TESS light curves from MAST \citep{https://doi.org/10.17909/fwdt-2x66}. 
We primarily used light curves with a 2-minute cadence. For EPIC~212036875~b, we also used the 10-minute cadence light curves from TESS sector 44 reduced by the MIT Quick-Look Pipeline \citep{Huang2020}.                                       

\begin{figure}[ht!]
\centering
\includegraphics[width=1.01\linewidth]{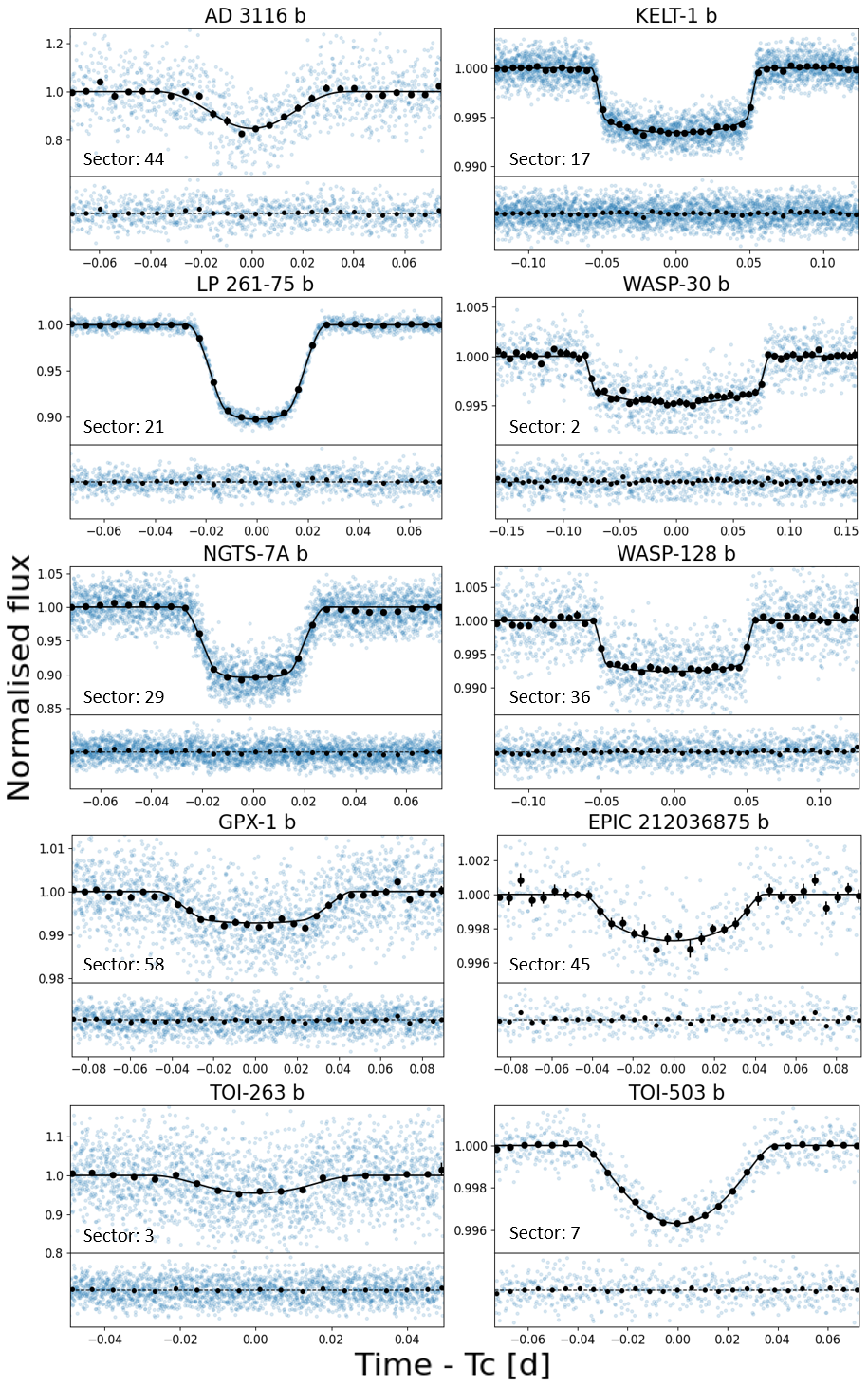}
\caption{ Phase-folded TESS transit light curves for all 10 BDs in our sample. Each light curve corresponding to data from a single TESS sector. The best-fit transit light curve model from \citet{Mandel2002} is shown in black solid line.}
\label{fig:PFLC}
\end{figure}

During our transit modeling, we used the \texttt{PyTransit} software \citep{Parviainen2015} to derive mid-transit times for each transit window for all BDs in our sample using the TESS light curves. 
For each transiting BD, we first performed an initial light curve fitting by applying the analytic transit model of \citet{Mandel2002} to each TESS sector data. We adopted the reference epoch $T_0$ and the orbital period \(P\) from the literature as priors, with the $T_0$ converted to $\mathrm{BJD_{TDB}}$ if necessary using the \href{https://astroutils.astronomy.osu.edu/time/hjd2bjd.html}{HJD to BJD} tool from \cite{Eastman2010}. 
Figure~\ref{fig:PFLC} shows the phase-folded transit light curves for all 10 BDs in our sample, with each light curve corresponding to data from a single TESS sector.
Next, we used the mid-transit time derived from this initial fit, adjusted by adding an integer multiple of the best-fit orbital period, as the updated prior for fitting each individual transit event within the corresponding sector.
This leads to a total of 319 TESS middle transit times for our sample of 10 BDs, each corresponding to an individual transit event from the TESS light curves (Table~\ref{tab:transit_time}).
The transit model parameters included the reference epoch $T_0$, orbital period $P$, impact parameter $b$, stellar density (in $\mathrm{g/cm^3}$), and the planet-to-star radius ratio $R_p/R_{\ast}$. The orbital eccentricity and argument of periastron were parameterized as $\sqrt{e} \cos\omega$ and $\sqrt{e} \sin \omega$, respectively. Additionally, two quadratic limb-darkening parameters were fitted. 

\begin{deluxetable*}{cccc}[ht]
\tablecaption{Transit Times. \label{tab:transit_time}}
\tablewidth{0pt}
\tablehead{
\colhead{System} & \colhead{Epoch} & \colhead{${T_{mid}}~$(BJD$_\mathrm{TDB}$)} & \colhead{Uncertainty (days)}
}
\startdata
AD 2116 b & 1171 & 2459500.67069 & 0.00205 \\
AD3116b & 1172 & 2459502.65226 & 0.00222 
\enddata
\tablecomments{Only a portion of this table is displayed here to illustrate its format and content. The complete, machine-readable version of the table can be accessed in the online version of this work.}
\end{deluxetable*}

\section{Timing Analysis} \label{sec:Timing}
In this section, we manually gathered as much data as possible from previous literature, along with the TESS transit times derived from this study, to update the linear ephemeris and search for potential long-term orbital decay in our sample of 10 BDs. However, many of our samples lacked follow-up observations after being discovered, relying solely on data from their discovery articles and TESS.

This work employed the Python software PdotQuest, developed in our previous work \citep{Wang24}, to model orbital period variations. The software supports both a linear model and a quadratic model. The linear ephemeris assumes a constant period model: 
\begin{equation}
T_N = T_0 +NP,
\end{equation}
In the linear model, the transit time for each epoch $T_N$ is calculated as a constant period $P$ from the reference epoch $T_0$(the time of the first transit).
The quadratic model accounts for a period change at a constant rate $\dot{P}$ \citep{Patra17}, expressed as 
\begin{eqnarray}
T_N &=& T_{N-1}+P_{N-1}, \\
P_N &=& P_{N-1}+\dot{P}(P_{N-1}+P_N)/2
\end{eqnarray}
In the quadratic model, $T_N$ is the predicted transit time at epoch $N$, and $P_N$ is the orbital period at epoch $N$, which changes due to tidal effects. 
Although tidal decay is inherently nonlinear over long astrophysical timescales, assuming a constant orbital period derivative over observational baselines spanning several tens of years is a reasonable approximation and is widely adopted in similar studies \citep{Patra17, Yee20, Yang22, Ma25}.
Further details about the software are available in \citet{Wang24}.

To compare the linear and quadratic models, we used the Bayesian Information Criterion (BIC), which helps identify the model that best balances goodness of fit with model complexity \citep{schwarz1978estimating}. It is calculated as:
\begin{equation}
BIC = \chi^2 + k \log n,
\end{equation}
where $n$ is the number of data points, $k$ is the number of free parameters in the model, and $\chi^2$ represents the goodness of fit between the model and the data. A lower BIC indicates a better model, and we used the difference in BIC values $\Delta BIC = BIC_{\text{lin}} - BIC_{\text{quad}}$ to select the most appropriate model. Specifically, a $\Delta BIC > 10$ is considered strong evidence in favor of the quadratic model \citep{Kass1995}.

We use two criteria to identify candidates with potential orbital period variations:
(a) the $\dot{P}$ is at least $3\sigma$ away from zero value, and (b) $\Delta BIC>10$.

To test the robustness of the quadratic model, we employed leave-one-out cross-validation (LOOCV) test to our long-term period variation candidates. This test involves removing one transit data point at a time and refitting the model to the remaining data. LOOCV helps identify the influential data points and assess the sensitivity of the model to the data, ensuring that the results are not overly reliant on any single observation.

\section{Results} \label{sec:Results}

Using the transit timing data from TESS and literature, we can put upper limits on the orbital decay (or expansion) rate of BDs, providing constraints on the evolution timescale of BD companions around solar-type stars. 
This way, we can test the different formation channels of BD companions around solar-type stars. 
Of the 10 BDs analyzed, five lacked sufficient data for a meaningful quadratic fit.
The remaining five transit BDs are discussed below, with their fitting parameters summarized in Table~\ref{tab:tab2}.

\begin{deluxetable*}{ccccccccc}[ht]
\tablecaption{Best-fit Transit Ephemerides for Transiting BDs \label{tab:tab2}}
\tablewidth{0pt}
\tablehead{
\colhead{System} & \colhead{Linear $P$(d)} & \colhead{Linear $T_0$} & \colhead{Quad. $P_0$(d)} & \colhead{Quad. $T_0$} &  \colhead{$\dot{P}$ (ms/yr)} & \colhead{BIC$_{linear}$} & \colhead{BIC$_{quad}$} & \colhead{$Q'_*$}
} 
\startdata
AD~3116~b & 1.98279436 & 2457178.81780 & 1.9827947 & 2457178.81778& -10.75 $\pm$ 13.36 & 42.82 & 45.90 & $2.2 \times 10^4$\\
&  $\pm$ 0.00000023 & $\pm$ 0.00007 & $\pm$ 0.0000005 & $\pm$ 0.00008 \\
KELT-1~b & 1.21749381 & 2455899.55342 & 1.21749426 & 2455899.55316 & -6.62 $\pm$ 2.03 & 155.13 & 148.59 & $6.1 \times 10^6$\\
&  $\pm$ 0.00000004 & $\pm$ 0.00008 & $\pm$ 0.00000015 & $\pm$ 0.00011 \\
LP~261-75~b & 1.88172235 & 2458159.73149 & 1.88172157 & 2458159.73151 & 35.75 $\pm$ 7.95 & 38.42 & 21.15 & $1.6 \times 10^4$\\
&  $\pm$ 0.00000005 & $\pm$ 0.00002 & $\pm$ 0.00000018 & $\pm$ 0.00002 \\
WASP-30~b & 4.1567786 & 2455334.98510 & 4.156777 & 2455334.98540 & 18.16 $\pm$ 24.14 & 26.32 & 28.92 & $8.8 \times 10^4$\\
&  $\pm$ 0.0000007 & $\pm$ 0.00070 & $\pm$ 0.000002 & $\pm$ 0.00080 \\
WASP-128~b & 2.20882284 & 2456720.68822 & 2.2088222 & 2456720.68823 & 5.49 $\pm$ 25.22 & 41.66 & 44.84 & $6.4 \times 10^5$\\
&  $\pm$ 0.0000002 & $\pm$ 0.00019 & $\pm$ 0.0000012 & $\pm$ 0.00019 \\
\hline
\colhead{System} & \multicolumn{2}{c}{Linear $P$(d)} & \multicolumn{2}{c}{Linear $T_0$} \\
\hline
EPIC 212036875 b & \multicolumn{2}{c}{5.1699000 $\pm$ 0.0000018} & \multicolumn{2}{c}{2458098.67911 $\pm$ 0.00020}\\
GPX-1 b & \multicolumn{2}{c}{1.7445772 $\pm$ 0.0000009} & \multicolumn{2}{c}{2458770.23820  $\pm$ 0.00040}\\ 
NGTS-7A & \multicolumn{2}{c}{0.67599085 $\pm$ 0.00000004} & \multicolumn{2}{c}{2457708.55410 $\pm$ 0.00001}\\
TOI-263 b & \multicolumn{2}{c}{0.5568083 $\pm$ 0.0000006} & \multicolumn{2}{c}{2458386.16970 $\pm$ 0.00060}\\ 
TOI-503 b & \multicolumn{2}{c}{3.6773554 $\pm$ 0.0000006} & \multicolumn{2}{c}{2458492.05297 $\pm$ 0.00019}\\
\enddata
\end{deluxetable*}

\subsection{AD~3116~b}

\begin{figure*}
\centering
\includegraphics[scale=0.4]{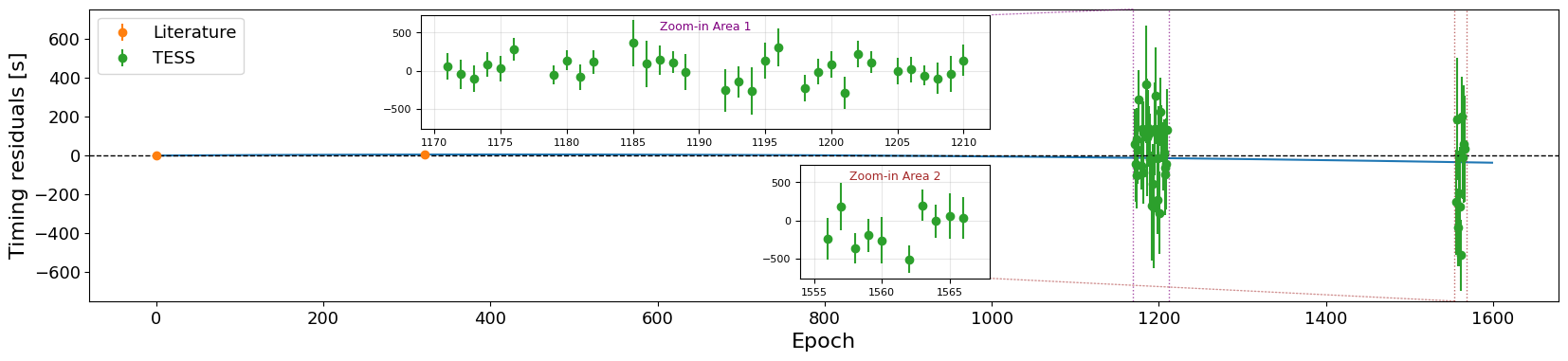}
\caption{Timing residuals of AD~3116~b.The blue curve indicates the best-fit quadratic model.
The orange points represent literature data, and the green points represent TESS data. We also over-plot zoom-in areas for some of the crowded TESS data points.} 
\label{fig:AD3116b}
\end{figure*}

AD~3116~b is a BD with a mass of 54.6~$M_J$ and a radius of 0.95~$R_J$, orbiting an M3.9-type dwarf star with an orbital period of 1.98~days \citep{Gillen2017}. 
AD~3116 is one of the few known M-dwarf systems with a transiting BD and is the youngest such system discovered in an open cluster, making it a valuable target for future research. 
In this study, we utilized two reference transit timing data from the literature: one from the discovery paper \citep{Gillen2017} and the other from \citet{Carmichael2023}, both are derived from multiple transit observations. 
Additionally, we have analyzed a total of 42 transits from TESS sectors 44, 45, 46, and 72. 
After performing the quadratic fitting using all these data, we do not detect any significant period change in the AD~3116 system, with the fitting results shown in Figure~\ref{fig:AD3116b}. The orbital period change rate $\dot{P}=-10.75 \pm 13.36$ ms/yr, implying an inspiral timescale ($\tau_{\rm inspiral} = P/\dot{P}$) of 15.9~Myr, and a $3\sigma$ lower limit of 3.4~Myr.

\subsection{KELT-1~b}
\begin{figure*}
\centering
\includegraphics[scale=0.4]{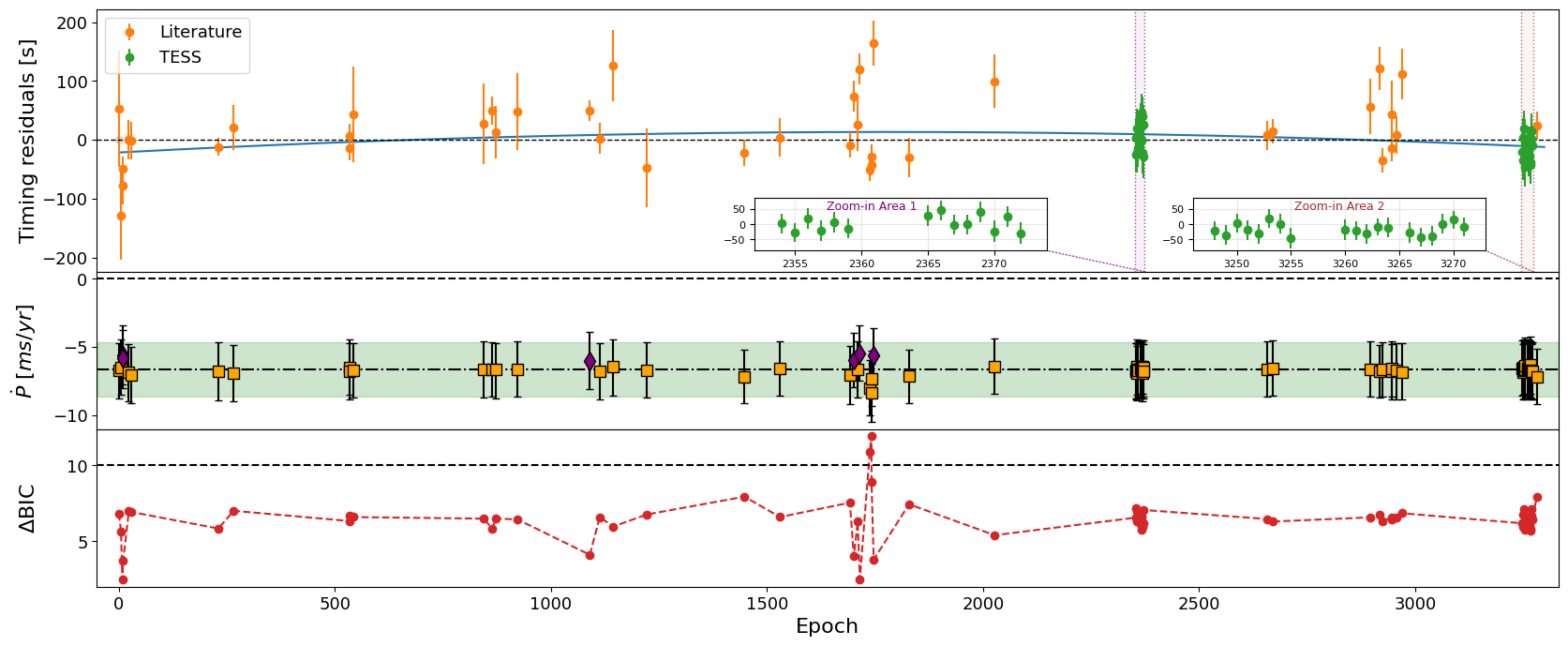}
\caption{Timing residuals and LOOCV analysis of KELT-1~b.
\textbf{Top panel}: Timing residuals from the TTV fitting. The blue curve indicates the best-fit quadratic model. The orange points represent literature data, and the green points represent TESS data. We also over-plot zoom-in areas for some of the crowded TESS data points.
\textbf{Middle panel}: LOOCV analysis showing the period change rate $\dot{P}$ after the removal of each single transit timing data point. The orange squares correspond to $\dot{P}$ values that are 3$\sigma$ away from zero, while the purple diamonds represent $\dot{P}$ values that do not meet this criterion. The dash-dotted line indicates the original best-fitting $\dot{P}$ before any data removal, and the green shaded area represents the 1$\sigma$ confidence region.
\textbf{Bottom panel}: Corresponding $\Delta$BIC values from the LOOCV analysis.}
\label{fig:KELT1b}
\end{figure*}

KELT-1~b, a BD with a mass of 27.23~$M_J$ and a radius of 1.11~$R_J$, orbits an F5-type dwarf star with an orbital period of 1.22~days \citep{Siverd12}. 
In this study, we utilized 41 archival data points from literature. One data point is taken from the original discovery paper \citep{Siverd12}, which is a composite data point derived from multiple transits. The other 40 data points are from \citet{Basturk2023}, including six from the Exoplanet Transit Database (ETD), 17 from various literature sources, and 17 from their own observations. 
We have analyzed a total of 33 transits of KELT-1~b from TESS sectors 17 and 57. 
By combining data from the literature with our TESS data, we performed a quadratic fit, yielding an orbital period decay rate of -6.62 $\pm$ 2.03~ms/yr for KELT-1~b. 
This implies an inspiral timescale of 15.9~Myr, with a $3\sigma$ lower limit of 8.3~Myr.
The $\dot{P}$ value is $3\sigma$ away from 0, but did not meet the $\Delta BIC >10$ criterion ($\Delta BIC = 6.54$). 
The LOOCV test results indicated that the removal of multiple points consistently led to a decrease in $\dot{P}$ and a significant reduction in the $\Delta BIC$ value, which is shown in Figure~\ref{fig:KELT1b}. 
Thus, we can conclude that KELT-1~b does not exhibit significant orbital period decay. 

\subsection{LP~261-75~b}
\begin{figure*}
\centering
\includegraphics[scale=0.4]{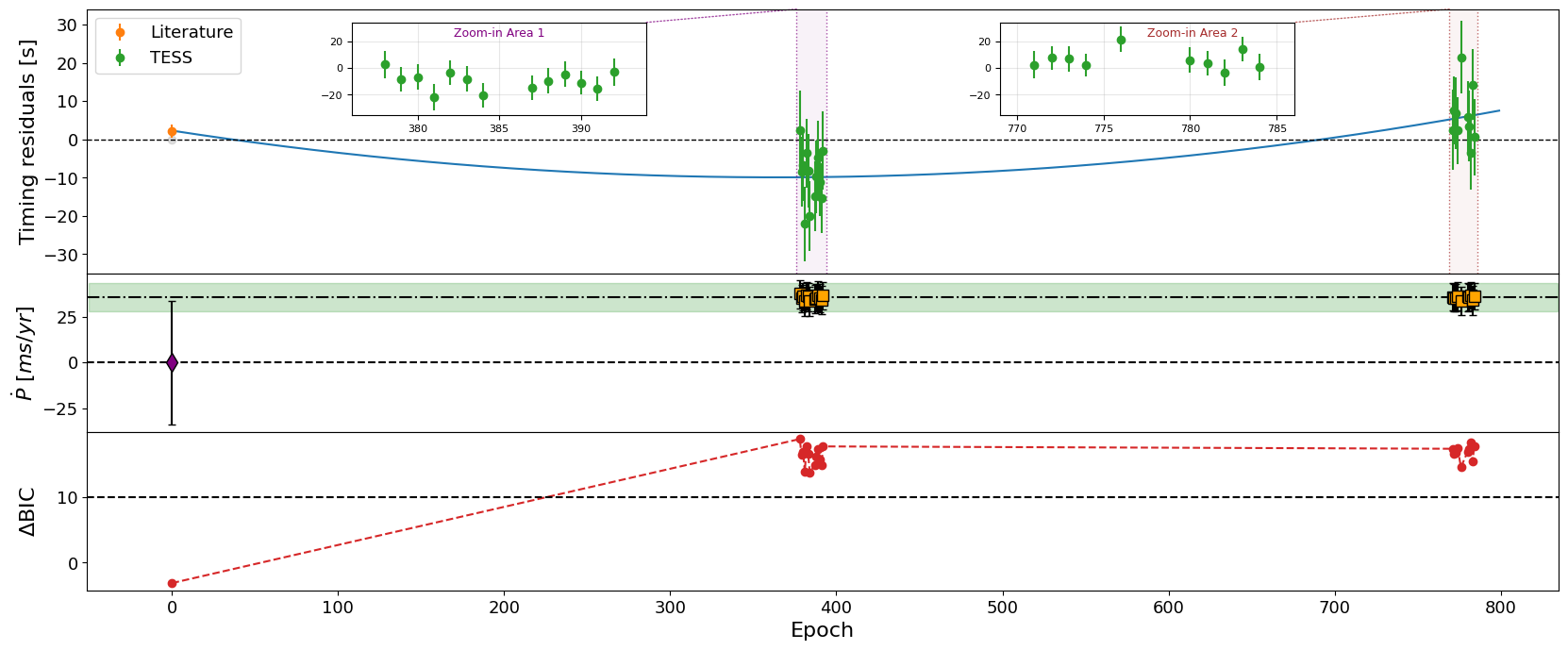}
\caption{Timing residuals and LOOCV analysis of LP 261-75~b. The lines and symbols are similar to those used in Figure~\ref{fig:KELT1b}.  } 
\label{fig:LP261_75b}
\end{figure*}

LP~261-75~b, also known as NLTT~22741~b, is a BD with a mass of 68.1~$M_J$ and a radius of 0.92~$R_J$ that orbits an active M4.5-type dwarf star with a period of 1.98~days \citep{Irwin2018}. 
The system was initially identified as a wide M4.5/L6 binary by direct imaging study of \citet{Reid2006}. 
Despite being discovered over a decade ago, LP~261-75~b has only a single transit timing measurement, from \citet{Irwin2018}. 
We analyzed a total of 23 transits from TESS sectors 21 and 48. 
After performing a quadratic fit, we found a positive period change rate of $\dot{P} = 35.75 \pm 7.95$~ms/yr. 
However, due to the short observational timespan and the availability of only one reference data, this system did not pass the LOOCV test, shown in Figure~\ref{fig:LP261_75b}. 
Additional data are required to verify this positive $\dot{P}$.

\subsection{WASP-30~b}
\begin{figure*}
\centering
\includegraphics[scale=0.4]{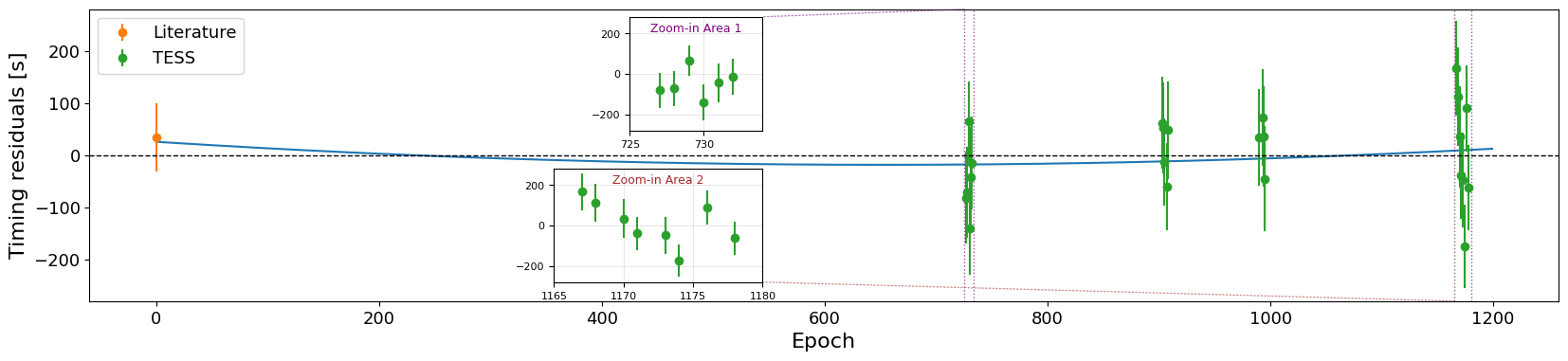}
\caption{Timing residuals of WASP-30~b.  The lines and symbols are similar to those used in Figure~\ref{fig:AD3116b}. } 
\label{fig:WASP30b}
\end{figure*}

WASP-30~b is a BD with a mass of 60.86~$M_J$ and a radius of 0.89 $R_J$, orbiting an F8V star with a period of 4.16 days. The system was discovered in 2010 \citep{Anderson11}. 
WASP-30 b also has only a single reference transit time \citep{Anderson11}.
We analyzed 23 transits from TESS sectors 2, 29, 42, 69, and 70. A quadratic fit revealed a positive trend in the period variation, though the associated error is too large for the result to be statistically significant (Figure~\ref{fig:WASP30b}). With a $\dot{P}$ of 18.16 $\pm$ 24.14 ms/yr, the $3\sigma$ lower limit on its inspiral timescale is 6.6~Myr.

\subsection{WASP-128~b}
\begin{figure*}
\centering
\includegraphics[scale=0.4]{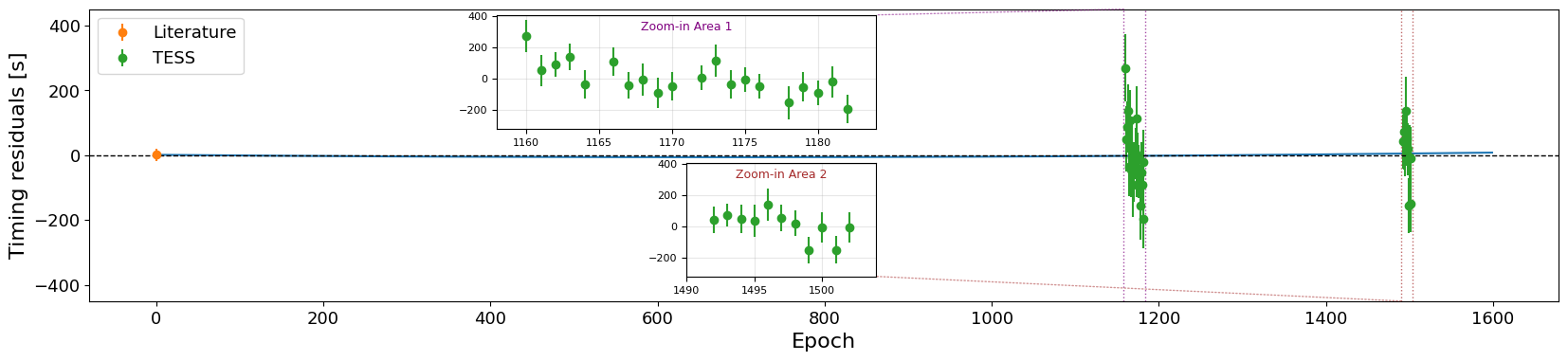}
\caption{Timing residuals of WASP-128~b. The lines and symbols are similar to those used in Figure~\ref{fig:AD3116b}.  } 
\label{fig:WASP128b}
\end{figure*}

WASP-128~b is a BD with a mass of 37.19~$M_J$ and a radius of 0.94~$R_J$, orbiting a G0V star with a period of 2.21 days. The system was discovered in 2018 \citep{Hodzic18}. WASP-128~b also has only one reference transit timing data \citep{Smith20}.
This data originates from the discovery paper \citep{Hodzic18}, which reported an incorrect orbital period. \citet{Smith20} reanalyzed the original light curves along with TESS long-cadence data to provide corrected values for $T_0$ and the orbital period, which we adopted in our analysis.
We analyzed 31 TESS transits from sectors 36, 37, and 63. A quadratic fit shows no significant orbital period variation in WASP-128~b with current data (Figure~\ref{fig:WASP128b}). 
With a $\dot{P}$ of 5.49 $\pm$ 25.22 ms/yr, the $3\sigma$ lower limit on its inspiral timescale can be constrained to 2.7~Myr.

\subsection{Other BDs}
For EPIC~212036875~b, GPX-1~b, and NGTS-7A~b, only a single transit timing data is available from their discovery papers. GPX-1~b and NGTS-7A~b each have TESS data from only one sector (Sector 58 and Sector 29, respectively). 
EPIC~212036875~b have TESS data from three sectors (Sector 45, 46 and 72), but the limited quantity and quality of the data prevent a meaningful quadratic fit. also lack sufficient coverage, with current TESS observations spanning too short a baseline for TTV analysis.
Despite these limitations, these systems—particularly the ultra-short-period TOI-263~b ($\sim$0.5 days)—have the potential to probe tidal orbital decay and warrant further observation.

\section{Discussion} \label{sec:Discussion}

\subsection{The Tidal Quality Factor}  \label{subsec:Q}
As suggested by \citet{Shahaf19}, several scenarios have been proposed to explain the formation of the BDD. One possibility is that BDs forming in the desert would be moved away due to strong tidal interactions with their F-, G-, and K-type host stars, causing them to either spiral inward or outward. To test this scenario, we can constrain the efficiency of these tidal interactions through the measurements of orbital period change rates of BDs within the BDD.

If we assume the orbital decay is entirely driven by tidal interactions, the modified tidal quality factor $Q'_*$ can be derived using Equation (20) of \cite{Goldreich1966}:
\begin{equation} \label{eq:Q}
Q'_*=-\frac{27}{2}\pi\biggl(\frac{M_p}{M_\star}\biggr)\biggl(\frac{a}{R_\star}\biggr)^{-5}\biggl(\frac{\mathrm{d}P}{\mathrm{d}t}\biggr)^{-1},
\end{equation}
where $M_p$ and $M_\star$ are the masses of the BD and its host star, $R_\star$ is the stellar radius, and $a$ is the orbital semi-major axis. 
The dimensionless parameter $Q'_*$ is used to describe the energy dissipation efficiency of the host star. It is sometimes expressed as $Q'_*=\frac{3Q_*}{2k_2}$, where $k_2$ is the Love number. 
It should be noted that Equation~(\ref{eq:Q}) is based on the `constant phase lag model' and is most applicable under the assumption that tidal dissipation is induced by equilibrium tides. 
Therefore, Equation~(\ref{eq:Q}) holds under the assumptions of a circular orbit, a stellar angular velocity much smaller than the orbital angular velocity, and negligible tidal dissipation within the BD.

For all the five BD systems, we use the nominal values of $\dot{P}$ to estimate their $Q'_*$. For the three BDs with positive $\dot{P}$ values, LP~261-75~b, WASP-30~b, and WASP-128~b, we calculate $Q'_*$ by dropping the negative sign in Equation~(\ref{eq:Q}).
Given the lack of strong statistical significance in the $\dot{P}$ values, the derived $Q'_*$ values should be treated with caution and regarded as preliminary until more precise observational data become available.
The physical parameters used to calculate $Q'_*$ for the BDs are summarized in Table~\ref{tab:tab3}. 

Next, we calculate the theoretical $Q'_*$ for BDs using the equilibrium tidal theory, which provides a framework for modeling tidal dissipation in such massive sub-stellar objects.
First, we calculate the potential orbital period variation rates (assumed to be negative) for our 10 BD systems. 
Following the work of \cite{Nordhaus13}, we evaluate the orbital period decay rate for each BD in our sample by
\begin{equation}\label{eq:pdot}
    \frac{\dot{P}}{P} = \frac{3}{2}\frac{\dot{a}}{a} = -\frac{18k_{2,*}f}{\tau_{*,conv}}\frac{M_*^{\rm env}}{M_*}\frac{M_{\rm BD}}{M_*}(1+\frac{M_{\rm BD}}{M_*})(\frac{R_*}{a})^8,
\end{equation}
where $k_{2,*}$ is the tidal Love number of the primary star, $f$ is a dimensionless parameter. In our calculation, we set $k_{2,*}$ and $f$ to be unity. 
The primary convective timescale is defined by ${\tau_{\rm *,conv} \equiv M_*^{\rm env}R_*^2/L_*}^{1/3}$, where $L_*$, $R_*$ and 
$M_*^{\rm env}$ are the luminosity, radius, and convective envelope mass of the primary star. We calculated these values using the Single-star Evolution (SSE) grid in the COMPAS code \citep{Riley22teamCOMPAS}.
Then, combining Equation~\ref{eq:Q} and \ref{eq:pdot}, we can derive the theoretical $Q'_*$ values for each BD system in our sample. 

\begin{figure*}
\centering
\includegraphics[scale=0.6]{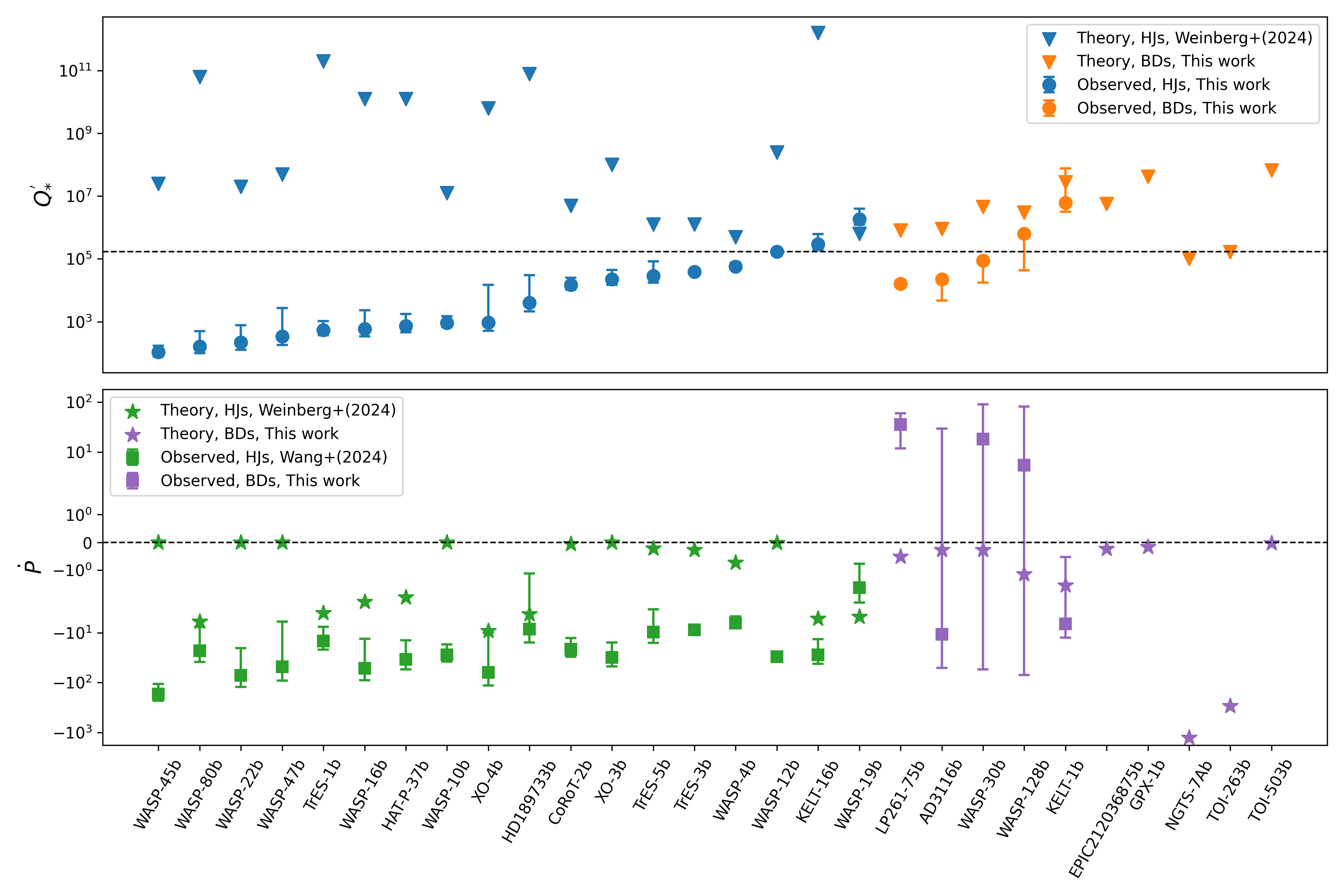}
\caption{Comparison of theoretical and observed values for $Q'_{*}$ and $\dot{P}$ for for HJs and BDs. The upper panel shows the comparison of $Q'_{*ob}$ and $Q'_{*th}$ for HJs and BDs. Blue downward triangles represent the theoretical values of $Q'_{*}$ values for HJs from \citet{Weinberg24}, while blue circles represent the observed $Q'_{*}$ values for HJs. Orange downward triangles show the theoretical $Q'_{*}$ values for BDs, and orange circles represent the observed $Q'_{*}$ values for BDs.
The lower panel compares the $\dot{P}_{ob}$ and $\dot{P}_{th}$ for HJs and BDs. Green stars represent the theoretical $\dot{P}$ values for HJs, calculated from the corresponding $\dot{P}$ using Equation~(\ref{eq:Q}) and green squares represent the observed $\dot{P}$ values for HJs from \citet{Wang24}. Purple stars represent the theoretical $\dot{P}$ values for BDs, and purple squares represent the observed $\dot{P}$ values for BDs.} 
\label{fig:Qob compare with Qth}
\end{figure*}

One interesting topic in the study of BDD, is to compare the properties of BDs in the desert and HJs \citep{Ma14}. 
Thus here we also compare the $Q'_*$ parameters between short period transiting BD systems and transiting HJ systems. 
We have selected 18 HJs from our previous work \citep{Wang24}, which show signs of tidal orbital decay. 
For these HJs, their observed $Q'_*$ are calculated using their measured $\dot{P}$ and Eqs.~\ref{eq:Q}, as listed in Table \ref{tab:tab4}. Their theoretical $Q'_*$ are estimated using the grids provided by \citet{Weinberg24}. 
The results of all theoretical $Q'_*$ and $Q'_*$ derived from $\dot{P}$ for transiting HJ and BD are shown in Figure~\ref{fig:Qob compare with Qth}.
We can see that the theoretical and observed $\dot{P}$ values of transiting BD and HJ generally fall within similar ranges, but for systems like LP~261-75~b and WASP~128~b, the observed values differ significantly. 
This again reminds us that, due to the limited data available for transiting BDs, the current $\dot{P}$ and $Q'_*$ results are not yet precise enough to support a robust comparison.
In this plot, we also observe that for hot Jupiters, the observed $\dot{P}$ generally do not agree well with the theoretical predictions made by \citet{Weinberg24}. 
For BD companions, the agreement between observed values and theoretical predictions appears better, likely due to the large error bars in the $\dot{P}$ measurements.
It is safe to conclude that there remains a large discrepancy between the orbital period decay rate measured using the TTV technique and that derived using tidal decay theory. 
There are several potential explanations. The first is that the accuracy of the orbital period decay measurements still has plenty of room to improve. Secondly, beside tides, other physical mechanisms can also cause the measured orbital period decay, including mass loss, Applegate effect, Romer effect, and a third body in the system, etc. 
Thus, most of the studies trying to constrain $Q'_*$ from observed $\dot{P}$ can only provide an effective lower limit.
And from Figure~\ref{fig:Qob compare with Qth}, we can see that the observed $Q'_*$ values do generally are lower than the theoretical values.
Thirdly, the theoretical calculation of orbital period decay caused by tides needs to improve, such as when to use the equilibrium tide model, when to use the dynamical tide model. 

\subsection{Future Observations}
We find most of the transiting BDs do not have enough transits observed to measure their $\dot{P}$ precisely. Thus future observations are needed if we want to study their orbital dynamics. Here we conduct a simulation to study what is the best observation cadence strategy in the near future to measure the $\dot{P}$ for the 10 transiting BD systems studied in this work. 

We first generate additional transit timing data points by adopting timing error bar similar to that of TESS observation. 
To make the simulated data more similar to real data, Gaussian noises are added to the predicted true values. 
The standard deviation ($\sigma$) of the Gaussian noise is set to match the median size of the TESS error bars for the same transiting BD target, and the mean ($\mu$) is set to 0.
Then we use the method described in Section~\ref{sec:Timing} to analyze the simulated timing data, together with the timing data collected in this study. 
Through our simulations, we found that the initial choice of $\dot{P}$ value in the simulation has little impact on the resulting uncertainty, $\sigma_{\dot{P}}$. 
Since the true $\dot{P}$ values for transiting BDs in our sample are unknown, we choose to set them to zero for simplicity. 

We have tested a total of 12 different sets of follow-up observation strategies, comprising combinations of four different follow-up periods (5, 10, 15, and 20~years) and three different observation cadences. 
The three different cadences include single transit observation at the end of the follow-up period, 8 transit observations evenly distributed across the follow-up period, and two clusters of four consecutive transit observations scheduled at the middle and end of the follow-up period. 
Given the limitations of observational resources, we have focused on using a small number of observation points while emphasizing on the extension of the time span.

\begin{figure*}
\centering
\includegraphics[scale=0.6]{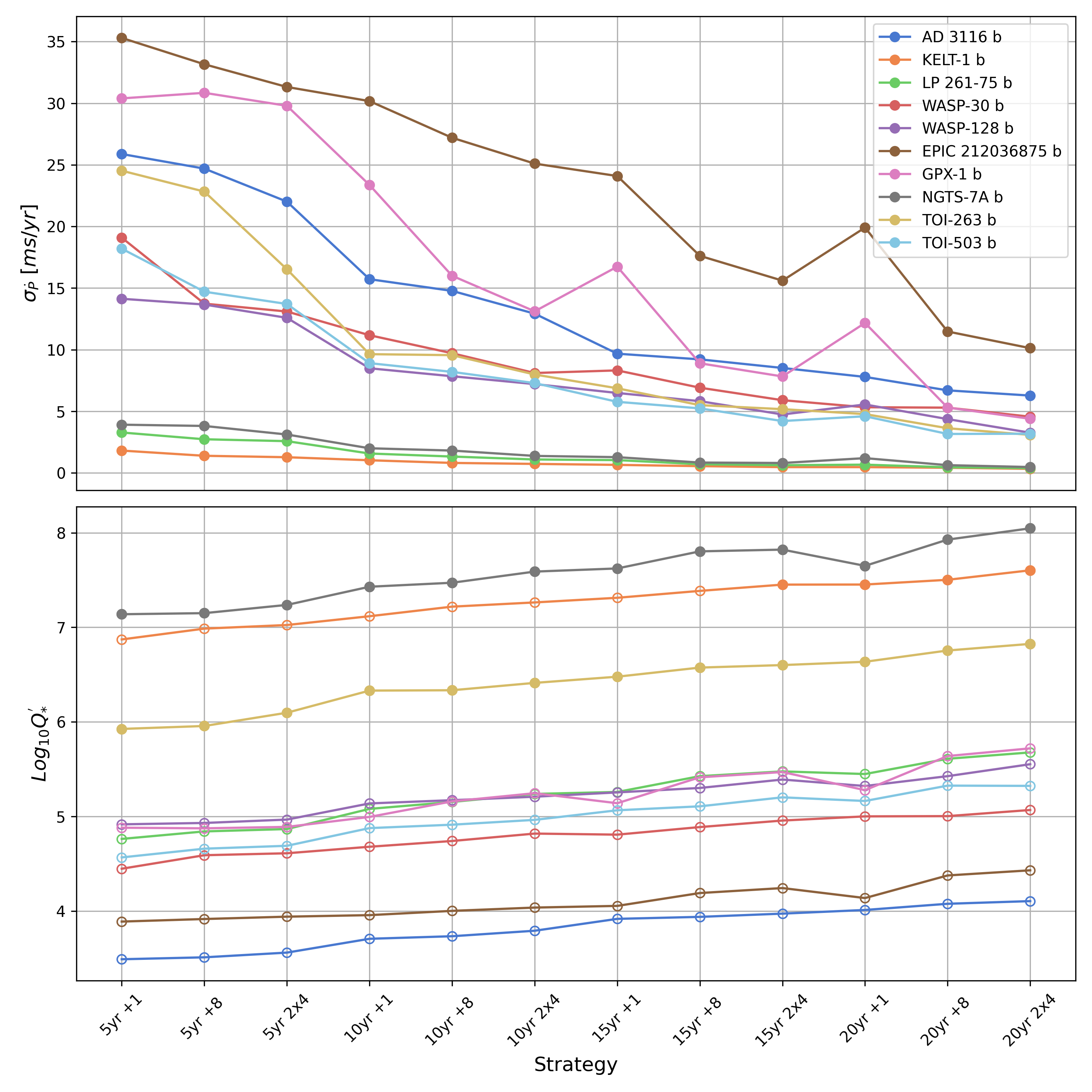}
\caption{Strategy assessment plot. In the top panel, we show the $\sigma_{\dot{P}}$ values that can be detected by using various observation strategy. The $Q'_*$ values corresponding the $3\sigma_{\dot{P}}$ are shown in the bottom panel. Solid points in the bottom panel indicate $Q'*$ values exceeding the theoretical value calculated in Subsection~\ref{subsec:Q}, while hollow points represent $Q'*$ values below the theoretical value. From this evaluation plot, we can see the longer the time baseline, the more observation data points obtained, the more easily to measure smaller $\dot{P}$ values.} 
\label{fig:Strategies}
\end{figure*}

The code used to generate and fit the simulated data has been updated in the PdotQuest package. For each observation strategy, the uncertainty $\sigma_{\dot{P}}$ in the period change rate $\dot{P}$ was derived from fitting the simulated data. 
As noted in Section~\ref{sec:Timing}, one of the criteria for classifying a system as exhibiting significant orbital decay is that $\dot{P}$ must be at least $3\sigma$ away from zero, i.e., $\dot{P} \leq -3\sigma_{\dot{P}}$. Using this minimum detectable $\dot{P}$, the corresponding tidal quality factor $Q'_*$ can be calculated via Equation~(5). 
The results, shown in Fig.~\ref{fig:Strategies}, illustrate how different follow-up observation strategies can constrain the minimum detectable $\dot{P}$ and tidal quality factor $Q'_*$ for each of the BD system. 
To ensure the minimum detectable $\dot{P}$ for the majority (9 out of 10) of BDs in our sample reaches the orbital period decay rate of WASP-12~b ($-30.19 \pm 0.92$~ms/yr; \citealt{Wang24}), at least 8 evenly spaced observations over a 15-year follow-up period (the `15yr+8' strategy in Fig.~\ref{fig:Strategies}) are needed.

Our simulations show that the uncertainty of the measured $\dot{P}$ depends not only on the number and precision of additional observations, but, more importantly, on the time span over which the new data are distributed.
Specifically, extending the follow-up period leads to a notable decrease in the $\sigma_{\dot{P}}$ values. 
We also found that, within the same time span, grouping transit observations into two clusters rather than distributing them evenly improves the accuracy of the $\dot{P}$ fit.
Therefore, we recommend that when researchers aim to obtain a more precise $\dot{P}$ for transiting BDs or HJs through future observations, they should consider scheduling several consecutive transit observations (i.e., `grouping') at intervals spaced by at least several years (`extending'). 
This optimal observing strategy, referred to as `grouping and extending', can help maximize the efficiency of observational resources while enhancing sensitivity to long-term orbital period variations.

Combining the observational strategy analysis with theoretical predictions of $\dot{P}$ from the previous subsection, we identify NGTS-7A~b and TOI-263~b as the most promising orbital decay candidates for future observations. 
Due to their extremely short orbital period (P = 16.22~hours and 13.36~hours), they are expected to undergo rapid tidal orbital decay ($>$200~ms/yr, as shown in Figure~\ref{fig:Qob compare with Qth}).
For NGTS-7A~b, given the high precision of TESS data, adding a single observation data point five years after the last recorded TESS observation (which occurred in 2020) should be sufficient to constrain its $\dot{P}$ value with enough precision to verify whether orbital period decay is present for this BD.
LP~261-75~b also warrants further study due to the high precision of its TESS transit data, adding a few transit timing data points in the next five years can verify its positive $\dot{P}$.

\begin{deluxetable*}{cccccccc}[ht]
\tablecaption{Physical parameters used to derive $Q'_*$ for HJs and BDs in this study. \label{tab:tab3}}
\tablewidth{0pt}
\tablehead{
\colhead{System} & \colhead{P(d)} &\colhead{$M_*(M_{\bigodot})$} & \colhead{$M_p~(M_J)$} & \colhead{$R_*(R_{\bigodot})$} & \colhead{$a/R_*$} & \colhead{Age(Gyr)} & \colhead{Ref} 
}
\startdata
AD~3116~b 
& 1.98279
& 0.28
& 54.20
& 0.29
& 16.00
& $<$1 
& (1)(2)\\
KELT-1~b 
& 1.21751
& 1.39
& 27.93
& 1.49
& 3.63
& 1.75 $\pm$ 0.25
& (1)(3)\\
LP~261-75~b
& 1.88172
& 0.30
& 68.08
& 0.31
& 13.84
& 0.15$\pm$0.05
& (4)\\
WASP-30~b 
& 4.15674
& 1.23
& 61.86
& 1.42
& 8.33
& 2.0$\pm$1.0
& (1)(5)\\
WASP-128~b 
& 2.20884
& 1.21
& 39.26
& 1.17
& 6.52
& 2.2$\pm$0.9
& (1)(6)\\
EPIC~212036875~b
& 5.16992
& 1.29
& 52.3
& 1.50
& 9.34
& 5.1$\pm$0.9
& (7)\\
GPX-1~b
& 1.74457
& 1.68
& 19.70
& 1.56
& 4.67
& $0.27^{+0.09}_{-0.15}$
& (a)\\
NGTS-7~A~b
& 0.67599
& 0.48
& 75.5
& 0.61
& 4.16
& $0.055^{+0.08}_{0.03}$
& (8)\\
TOI-263~b
& 0.55681
& 0.44
& 61.6
& 0.44
& 4.92
& 4.75$\pm$4.35
& (9)\\
TOI-503~b
& 3.67720
& 1.80
& 53.7
& 1.70
& 7.17
& $0.18^{+0.17}_{-0.11}$
& (10)\\
CoRoT-2~b
& 1.74299
& 0.97 
& 3.31 
& 0.90
& 6.71 
& $2.7^{+3.2}_{-2.7}$ 
& (a)\\
HD189733~b 
& 2.21857
& 0.83
& 1.16
& 0.78
& 8.92
& $6.80^{+5.20}_{-4.40}$
& (a)\\
HAT-P-37~b
& 2.79744
& 0.93
& 1.19
& 0.88
& 9.19
& $3.6^{+4.1}_{-2.2}$
& (a)\\
KELT-16~b 
& 0.96899
& 1.21
& 2.75
& 1.36
& 3.23
& 3.1$\pm$0.3
& (a)\\
TrES-1~b 
& 3.03007
& 1.04
& 0.84
& 0.85
& 10.52
& $3.7^{+3.4}_{-2.8}$
& (a)\\
TrES-3~b 
& 1.30619
& 0.93
& 1.91
& 0.83
& 6.0
& $0.90^{+2.8}_{-0.8}$
& (a)\\
TrES-5~b
& 1.48225
& 0.90
& 1.80
& 0.86
& 6.19
& 7.4$\pm$1.9
& (a)\\
WASP-4~b 
& 1.33823
& 0.86
& 1.18
& 0.89
& 5.45
& 7.0$\pm$2.9
& (a)\\
WASP-10~b 
& 3.09273
& 0.75
& 3.21
& 0.70
& 11.65
& $7.0^{+6.0}_{-3.0}$
& (a)\\
WASP-12~b 
& 1.09142
& 1.35
& 1.41
& 1.57
& 3.04
& $2.0^{+0.7}_{-2.0}$
& (a)\\
WASP-16~b 
& 3.11860
& 1.02
& 0.91
& 1.14
& 8.21
& $2.3^{+5.8}_{-2.2}$
& (a)\\
WASP-19~b 
& 0.788839
& 0.97
& 1.15
& 1.00
& 3.46
& $6.4^{+4.1}_{-3.5}$
& (a)\\
WASP-22~b 
& 3.53269
& 1.11
& 0.57
& 1.22
& 8.38
& $4.3^{+1.6}_{-1.1}$
& (a)\\
WASP-45~b 
& 3.12608
& 0.91
& 1.01
& 0.95
& 9.50
& $12.7^{+1.0}_{-5.3}$
& (a)\\
WASP-47~b 
& 4.15915
& 1.04
& 1.14
& 1.14
& 9.70
& $6.5^{+2.6}_{-1.2}$
& (a)\\
WASP-80~b 
& 3.06785
& 0.57
& 0.54
& 0.57
& 12.63
& 7.0$\pm$7.0
& (a)\\
XO-3~b 
& 3.19152
& 1.21
& 11.70
& 1.38
& 7.05
& $2.82^{+0.58}_{-0.82}$
& (a)\\
XO-4~b 
& 4.12508
& 1.32
& 1.61
& 1.56
& 7.68
& 2.10$\pm$0.60
& (a)\\
\enddata
\tablecomments{References: (1) \cite{Carmichael2023};(2) \cite{Gillen2017};
(3) \cite{Beatty2014}
(4) \cite{Irwin2018};
(5) \cite{Anderson11}
(6) \cite{Hodzic18}
(7) \cite{Carmichael2019};
(8) \cite{Jackman2019};
(9) \cite{Palle21};
(10) \cite{Subjak2020}; 
(a) \href{https://exoplanetarchive.ipac.caltech.edu/index.html}{NASA Exoplanet Archive}}
\end{deluxetable*}

\begin{deluxetable*}{cccc}[ht]
\tablecaption{Tidal quality  factor estimation for HJs \label{tab:tab4}}
\tablewidth{0pt}
\tablehead{
\colhead{System} & \colhead{$\dot{P}$ (ms/yr)} & \colhead{$Q'_*$} & \colhead{$\tau_{\rm inspiral}$~(Myr)}
}
\startdata
CoRoT-2 b & -21.65 $\pm$ 2.96 & $>1.5\times 10^4$ & 7.0\\
HD189733 b & -8.44 $\pm$ 2.44 & $>4.0\times 10^3$ & 22.7\\
HAT-P-37 b & -34.29 $\pm$ 6.67 & $>7.3\times 10^2$ & 7.0\\
KELT-16 b & -27.82 $\pm$ 4.76 & $>3.0\times 10^5$ & 3.0\\
TrES-1 b & -14.82 $\pm$ 2.37 & $>5.4\times 10^2$ & 17.7\\
TrES-3 b & -8.77 $\pm$ 0.47 & $>3.8\times 10^4$ & 12.9\\
TrES-5 b & -9.71 $\pm$ 2.11 & $>2.9\times 10^4$ & 13.2\\
WASP-4 b & -6.43 $\pm$ 0.56 & $>5.7\times 10^4$ & 18.0\\
WASP-10 b & -27.74 $\pm$ 3.49 & $>9.2\times 10^2$ & 9.6\\
WASP-12 b & -30.19 $\pm$ 0.92 & $>1.7\times 10^5$ & 3.1\\
WASP-16 b & -51.98 $\pm$ 12.87 & $>5.9\times 10^2$ & 5.2\\
WASP-19 b & -1.64 $\pm$ 0.29 & $>1.9\times 10^6$ & 41.6\\
WASP-22 b & -71.76 $\pm$ 17.12 & $>2.2\times 10^2$ & 4.3\\
WASP-45 b & -169.21 $\pm$ 21.09 & $>1.1\times 10^2$ & 1.6\\
WASP-47 b & -48.45 $\pm$ 14.14 & $>3.4\times 10^2$ & 7.4\\
WASP-80 b & -23.04 $\pm$ 5.15 & $>1.6\times 10^2$ & 11.5\\
XO-3 b & -31.59 $\pm$ 5.24 & $>2.2\times 10^4$ & 8.7\\
XO-4 b & -62.57 $\pm$ 17.48 & $>9.2\times 10^2$ & 5.7\\
\enddata
\end{deluxetable*}

\section{Conclusion}

In this study, we analyzed the long-term TTVs of 10 transiting BD systems with short orbital periods using data from TESS and previous studies. 
We find that five BD systems show potential period changes when fitting with a quadratic model.
However, these signals are not statistically significant.
Our simulations of future observations reveal that longer time spans are crucial for constraining $\dot{P}$, identifying NGTS-7A b, TOI-263~b and LP 261-75 b as the most promising candidates for follow-up observations. 

Our study demonstrates the potential power of long-term TTV monitoring in measuring the orbital decay rates of transiting BD systems. Future observation should prioritize continuous TTV monitoring of transiting BD systems, especially those with short orbital periods.
The orbital decay rates can place constraints on the survival timescale of BDs and may help refine models of the BDD around solar-type stars. 
Thus similar studies in the future can verify the orbital decay rates measured here, and have significant implications for understanding BD formation and evolution around FGK-type stars. 

~\\
We acknowledge the financial support from the National Key R\&D Program of China (2020YFC2201400), NSFC grant 12073092, 12103097, 12103098, the science research grants from the China Manned Space Project (No. CMS-CSST-2021-B09), and the Earth 2.0 research funding from SHAO.


\vspace{5mm}
\facilities{TESS}

\software{pytransit \citep{2015MNRAS.450.3233P}, PdotQuest \citep{Wang24}
}

\bibliography{sample631}
\bibliographystyle{aasjournal}

\end{document}